\documentclass[amsmath,amssymb,aps,reprint,prb,superscriptaddress]{revtex4-2}

\usepackage{graphicx} 
\usepackage{newtxtext,newtxmath}
\usepackage{physics}
\usepackage{svg}
\usepackage[T1]{fontenc}
\usepackage{bm} 
\usepackage{hyperref}
\hypersetup{colorlinks,linkcolor={blue},citecolor={blue},urlcolor={blue}}
\usepackage{upgreek}

\begin{document}

\title{
\textbf{Ferroic Polarization from Nonpolar Phonons}
}

\author{Seongjoo Jung}
\affiliation{Department of Chemical Engineering and Materials Science, University of Minnesota, Minneapolis, Minnesota 55455, USA}
\affiliation{Department of Materials, ETH Z\"{u}rich, CH-8093, Switzerland}
\author{Turan Birol}
\email{Corresponding author: tbirol@umn.edu}
\affiliation{Department of Chemical Engineering and Materials Science, University of Minnesota, Minneapolis, Minnesota 55455, USA}

\begin{abstract}
Born effective charge, a fundamental quantity in lattice dynamics and ferroelectrics, provides a quantitative measure of linear polarization response to ionic displacements. However, it does not account for higher-order effects, which can play a significant role in certain materials, such as fluorite HfO$_2$. In this paper, we use the second-order mode effective charges defined with the second-order atomic dynamical charges as a measure of the dipole moments generated by nonpolar lattice distortions. Using first-principles calculations, we demonstrate that specific combinations of nonpolar phonons in many oxides can induce strongly aligned second-order polarizations, reaching magnitudes comparable to those of intrinsically polar modes even in the zero frequency limit, broadening the understanding of second-order effects, which have historically been emphasized for their dynamical effects at specific frequency ranges. Through a symmetry-based analysis of the charge density, we elucidate the microscopic origin of these effects, tracing them to variations in bond covalency and local electronic rearrangements. We also demonstrate large second-order mode effective charge in well-studied perovskites, highlighting the generality of these phenomena. 
Our results offer insights into the design principles of next-generation ferroelectric, piezoelectric and multifunctional materials from the higher-order contribution to polarization in crystalline solids.
\end{abstract}

\maketitle

\section{Introduction}

The concept of dynamical charges, in particular the Born effective charge (BEC) $Z^*$---defined as the derivative of the macroscopic polarization with respect to ionic displacements---has played a critical role in advancing the understanding of Coulombic interactions in lattice dynamics, dielectric response, and ferroelectricity. This idea was first introduced in the pioneering works of Born, Göppert-Mayer, and Huang \cite{born1933dynamische,born1954dynamical}. Unlike static charges \cite{cochran1961effective}, the dynamical charge incorporates contributions from the reorganization of the electronic structure induced by ionic displacements \cite{ghosez1998dynamical,lichtensteiger2012ferroelectricity}. Anomalous BECs signal substantial modification of the bonding environment, such as covalent hybridization, when an ion is displaced. These anomalies are connected to important physical phenomena, including enhancement of static dielectric constant, ferroelectricity \cite{wang1996polarization}, large LO-TO splitting \cite{zhong1994giant}, and strong electron-phonon coupling \cite{verdi2015frohlich}.  With advances in first-principles methods such as density functional theory and the development of the modern theory of polarization \cite{hohenberg1964inhomogeneous,resta1992theory,king1993theory}, it has become possible to theoretically predict BECs in crystals \cite{resta1993towards,ghosez1994first,zhong1994giant}. In many cases, the polarization $P$ along $\alpha$ cartesian axis can be well approximated by a linear relation involving BECs and ionic displacements, $\Omega P_\alpha = \sum_{\kappa \beta} Z^{*}_{\alpha,\kappa\beta}u_{\kappa\beta}$ \footnote{The Born effective charge is a mixed derivative, and its notational convention varies across the literature. To accommodate straightforward expansion to higher-order terms, we adopt the specific notation $Z^*_{\alpha,\kappa\beta}=\Omega\pdv*{P_\alpha}{u_{\kappa\beta}}$. Also see Refs.~\cite{ghosez1998dynamical, gonze1997dynamical}.} where $\Omega$ represents the unit cell volume and $u_{\kappa\beta}$ the displacement of sublattice of ions $\kappa$ in direction $\beta$.

Nevertheless, the use of the BEC is inherently limited by its definition as a linear-response quantity. Variations in $Z^*$ as a function of atomic positions \cite{keating1965higher, keating1965first} are known to influence two-phonon infrared absorption \cite{deinzer2004two, fluckey2024diamond, born1954dynamical} and nonlinear susceptibility tensor in noncentrosymmetric compounds \cite{flytzanis1972infrared, roman2006ab}. 
Also, dependence of $Z^*$ on an external electric field is related to the Raman polarizability \cite{veithen2005,zhang2025rascbec}. While these variations of $Z^*$ and the second-order dipole moments they give rise to are often negligible in the static limit, they can be responsible of the dominant contributions in higher frequencies where the linear contributions vanish, especially in highly covalent or homopolar compounds \cite{born1954dynamical, keating1965first,keating1965higher}. To the best of our knowledge, the impact of the second-order dipole moment to the ferroic polarization in ferroelectric materials, especially in highly ionic systems such as perovskites, has not been studied in detail. (The word \textit{``ferroic''} refers to the spontaneous and switchable polarization in ferroelectric systems, which can be either proper or improper ferroelectrics. This is as opposed to e.g. polarization that may emerge under an external electric field in an otherwise nonpolar system.) We have recently shown that in HfO$_2$, higher-order contributions to dipoles associated with nonpolar modes can, in certain combinations, generate polarization comparable in magnitude to that arising from the polar mode itself \cite{jung2026triggered}. 
This significant nonlinear effect drives the variations in BECs across the different phases of HfO$_2$ \cite{fan2022vibrational}. 

\begin{figure*}[t]
\includegraphics[width=0.93\textwidth]{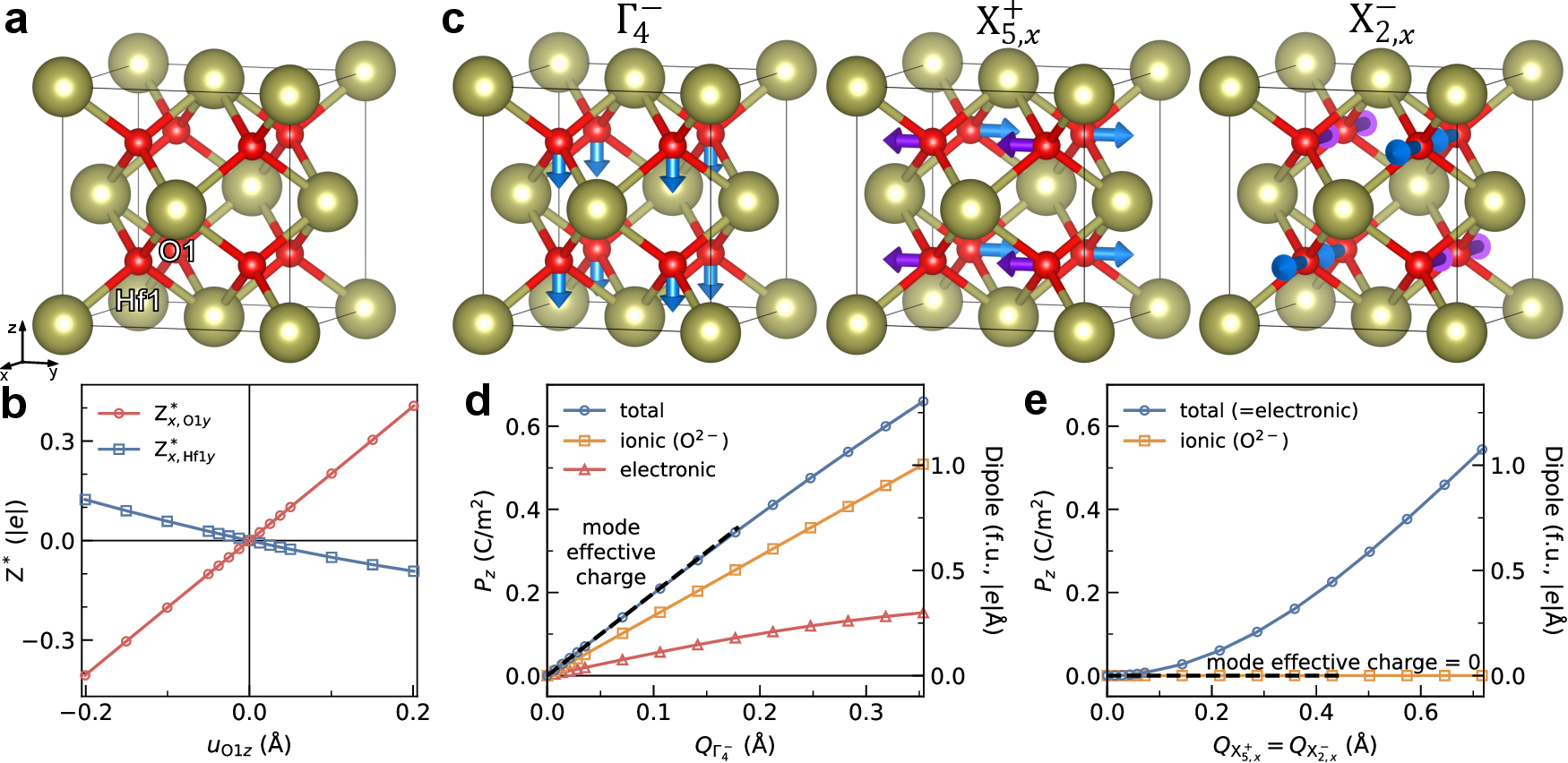}
\caption{\label{fig:high_dyn} 
    Second-order dipole moment in HfO$_2$.
    (a) Illustration of relative positions of O1 and Hf1 ion.
    (b) Change of Born effective charge components by ion displacement in cubic HfO$_2$. For a single oxygen displacement in the $z$ direction, largest changes are observed for $xy$ components for the Born effective charge of itself and adjacent Hf ions.
    (c) Illustration of $\Gamma^-_{4}, \mathrm{X}^+_{5,x}$ and $\mathrm{X}^-_{2,x}$ modes in HfO$_2$.
    (d) Linear interpolation of polar structure resulting from mode effective charge of polar $\Gamma^-_{4}$ mode in cubic HfO$_2$. 
    (e) Linear interpolation of polar structure resulting from 2nd-order mode effective charge of nonpolar modes $\mathrm{X}^+_{5,x}$ and $\mathrm{X}^-_{2,x}$  in HfO$_2$. Significant polarization can be induced from higher-order contribution of nonpolar modes. Note that the order parameters for the ground-state Pca$2_1$ HfO$_2$ is $0.40$, $0.40$, and $-0.48$ Å for $Q_{\Gamma_4^-}$, $Q_{\mathrm{X}_5^+}$, and $Q_{\mathrm{X}_2^-}$, respectively.}
\end{figure*}

In this paper, we introduce the concept of second-order \textit{mode} effective charge defined from the second-order dipole moments, and use first-principles calculations to elucidate their significant contributions in fluorite crystals. To develop physical insight into their origin, we employ group theoretical tools to project charge densities onto different irreducible representations (irreps) and atomic multipoles, which enables addressing key questions such as why the enhanced contributions to second-order dipole moments emerge specifically in the fluorite structure, but not in other systems where they are also allowed by symmetry, such as A$_3$B$_2$O$_7$ Ruddlesden--Popper phases. Finally, we show that the significant contributions from combinations of nonpolar phonons to the dynamical charge are not exclusive to fluorite structures, by identifying a phonon pair with high second-order mode effective charge in the cubic perovskite systems.

\section{Second-order dynamical charges and mode effective charges}

The second-order dynamical charge, $Z^{*(2)}$, is defined as a rank-3 tensor associated with a pair of sublattices, $\kappa$ and $\kappa'$:
\begin{equation}
Z^{*(2)}_{\alpha,\kappa\beta,\kappa'\gamma}
= \Omega \pdv{P_\alpha}{u_{\kappa\beta}}{u_{\kappa'\gamma}}
= \pdv{Z^{*}_{\alpha,\kappa\beta}}{u_{\kappa'\gamma}}
\label{def:2nd_dyn}
\end{equation}
This definition is similar to earlier formulations of ``second-order dipole moments,'' for instance as discussed in the context of two-phonon optical absorption spectra \cite{deinzer2004two}. In those settings, however, such quantities become relevant primarily when first-order contributions vanish, either by symmetry or because of an energy/frequency mismatch. 
Furthermore, in the context of static polarization in ferroelectrics, we refrain from using the term ``second-order dipole moment'' in order to avoid confusion with $\bm{p} = \Omega \bm{P}$ (Eq.~(\ref{eq:2nd_mode})). As with interatomic force constants, the second-order dynamical charges are pairwise quantities that can be evaluated either using density functional perturbation theory 
\footnote{Ref.~\cite{veithen2005} demonstrates a formalism for $\partial^3 E / \partial \mathcal{E} \partial \mathcal{E} \partial \mathcal{E}$ and $\partial^3 E / \partial \mathcal{E} \partial \mathcal{E} \partial u$ based on the $2n+1$ theorem. A similar formalism can be applied to $\partial^3 E / \partial \mathcal{E} \partial u \partial u$ to calculate second-order dynamical charges.}
or finite differences in a sufficiently large supercell by using Berry phase formalism (Supplementary Material S4). Note that the second-order dynamical charge is distinct from the dynamical quadrupole  \cite{martin1972piezoelectricity,royo2019first}, which also been found to exhibit a significant magnitude in HfO$_2$ \cite{macheda2024first}.

We define the second-order mode effective charge of modes $\lambda$ and $\mu$, which we denote as $\bm{Z}^{*(2)}_{\lambda\mu}$, as:
\begin{equation}
Z^{*(2)}_{\alpha,\lambda\mu}=\frac{\Sigma_{\kappa'\gamma}\Sigma_{\kappa\beta} Z^{*(2)}_{\alpha,\kappa\beta,\kappa'\gamma}U_{\lambda,\kappa\beta}U_{\mu,\kappa'\gamma}}{|\bm{U}_\lambda||\bm{U}_\mu|} \label{def:2nd_mode}
\end{equation}
where $\bm{U}_\lambda$ represents real-space displacements of mode $\lambda$ defined in the supercell \cite{gonze1997dynamical}. The information of the wavevector is implicit within the definition of $\bm{U}$. The $\lambda$ and $\mu$ modes don't need to be polar, or even zone-center modes. In general, the wavevectors of them need to add up to a reciprocal lattice vector for the product of their representations to include the polar $\Gamma$-point irrep and hence higher-order mode effective charge to be nonzero. The second-order contribution to dipole moment $\bm{p}$ from two nonpolar modes $\lambda$ and $\mu$ can be calculated using $Z^{*(2)}_{\alpha,\lambda\mu}$ as (Supplementary Material S2):
\begin{equation}
p_\alpha = NZ^{*(2)}_{\alpha,\lambda\mu}Q_\lambda Q_\mu \label{eq:2nd_mode}
\end{equation}
Where $N$ is the number of primitive cells in the supercell and $Q_\lambda$ is the order parameter for mode $\lambda$.

In Fig.~\ref{fig:high_dyn}(a--b), we show the two largest components of the second-order dynamical charges arising from an oxygen displacement in cubic HfO$_2$ (Fm$\bar{3}$m, \#225), namely $Z^{*(2)}_{x,\mathrm{O1}y,\mathrm{O1}z}$ and $Z^{*(2)}_{x,\mathrm{Hf1}y,\mathrm{O1}z}$ with respective values of 2.01, --0.54 $|e|/$Å, calculated from first principles. Panel (a) depicts the relative arrangement of the labeled ions. For cubic HfO$_2$, the off-diagonal components of the BEC tensor are zero. Panel (b) clearly shows the dependence of the (first-order) BEC on ionic displacements. The slope of the BEC near zero displacement defines the second-order dynamical charge.

An effect that stands out is that the dominant contributions to the second-order dynamical charge involve all three Cartesian indices \cite{roman2006ab}. When an oxygen ion is displaced along two different cartesian axes, a large polarization is induced along the direction normal to both displacements, as seen in the red data in Fig.~\ref{fig:high_dyn}(b). This translates into sizable second-order mode effective charges associated with ``hybrid modes'' (combinations of two or more modes belonging to different irreps) that involve $\mathrm{X}^+_{5,\alpha}$ and $\mathrm{X}^-_{2,\alpha}$, as well as between $\mathrm{X}^+_{5,\alpha'}$ and $\mathrm{X}^-_{5,\alpha':O}$ of cubic HfO$_2$, following the naming convention from Ref. \cite{jung2026triggered}. Note that the Cartesian indices associated with the modes are from their corresponding wavevectors, not the directions of the ionic displacements. This counterintuitive behavior implies that the majority of second-order dynamical charge contributions in HfO$_2$ stem from ionic motions perpendicular to the polarization axis, which complicates the intuitive connection between polarization and ionic displacements.

\begin{figure}[t]
\includegraphics[width=0.45\textwidth]{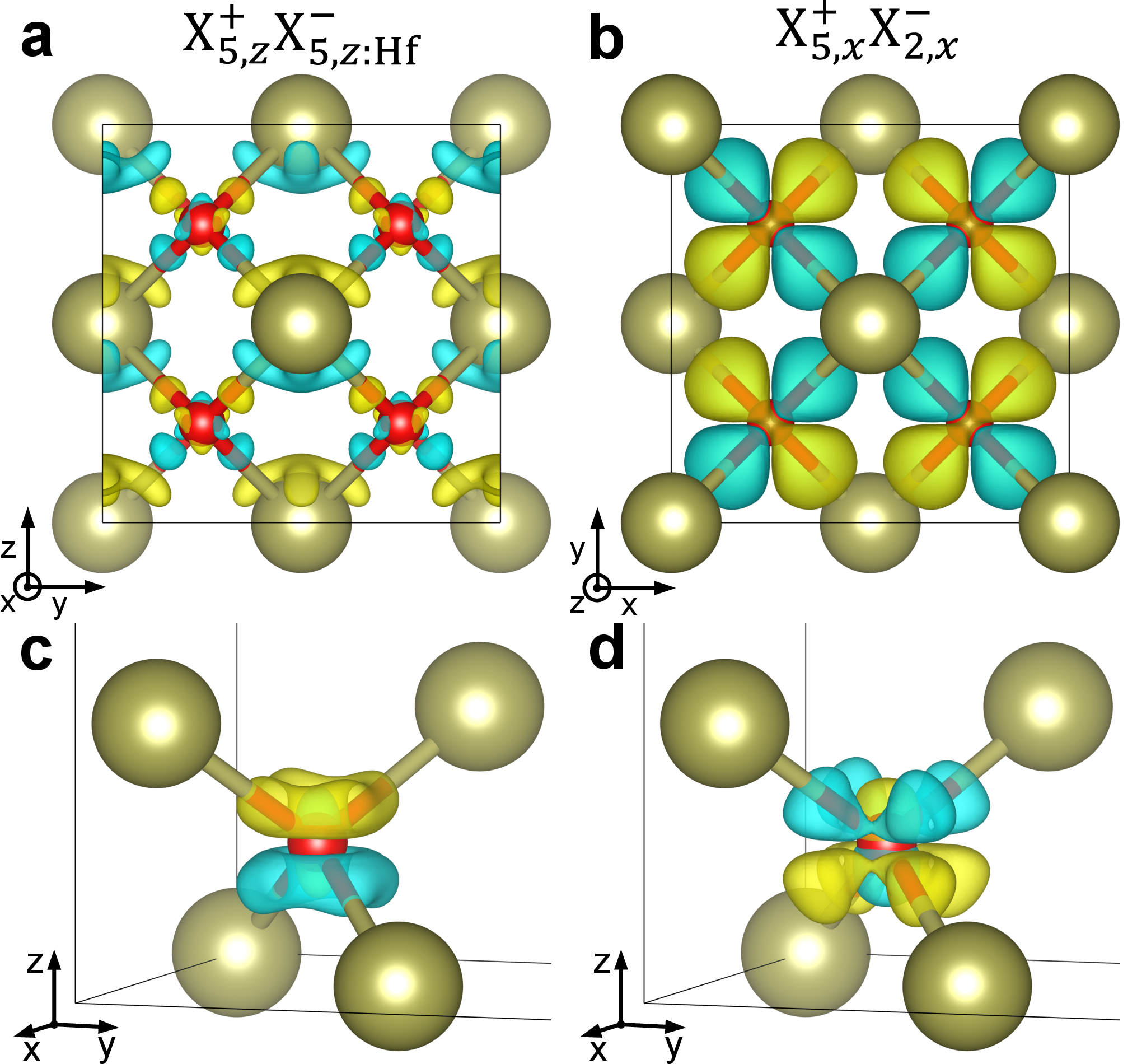}
\caption{\label{fig:charge_projection} 
    Isosurfaces of electron density projections of (a) $\mathrm{X}^+_{5,z}\mathrm{X}^-_{5,z:\mathrm{Hf}}$ (b) $\mathrm{X}^+_{5,x}\mathrm{X}^-_{2,x}$ hybrid nonpolar mode pairs onto the $\Gamma^-_4$ irrep. 
    Panels (c) and (d) show the inversion-odd components of the projected charge density on oxygen sites for the $\mathrm{X}^+_{5,z}\mathrm{X}^-_{5,z:\mathrm{Hf}}$ and $\mathrm{X}^+_{5,x}\mathrm{X}^-_{2,x}$ hybrid modes, respectively. 
    In all plots, yellow and cyan isosurfaces represent regions of electron (negative charge) accumulation and depletion. 
    The isosurface values are set to 12, 35 $\upmu|e|$/Å$^3$ for panels (a), (b), and 3.5 $\upmu|e|$/Å$^3$ for panels (c), (d). 
}
\end{figure}

The results we obtain by interpolating between the polar structures and the nonpolar parent structure, shown in Fig.~\ref{fig:high_dyn} (c--e), provide direct evidence for large higher-order dynamical charges. Panel (d) shows $z$ component of polarization induced by polar mode $\Gamma^-_{4}$ towards $z$, calculated by repeating Berry phase calculations at different structures with increasing amplitude of the polar mode. There are significant contributions from both ionic (static) and electronic (dynamical) components of the charge, where the ionic part is obtained by multiplying displacements with the nominal charges of +4 for Hf and --2 for O. In this case, the slope of the total $P_z$ is nonzero, reflecting a non-zero first-order mode effective charge associated with the polar distortion. In contrast, panel (e) shows $P_z$ exhibiting quadratic behavior when the amplitudes of two nonpolar modes $\mathrm{X}^+_{5,x}$ and $\mathrm{X}^-_{2,x}$ are increased together. This polarization has two distinctive features compared to panel (d): First, the initial slope of the polarization curve is zero, indicating that the polarization is generated exclusively through higher-order (nonlinear) mechanisms. Importantly, the interpolated curve exhibits no discontinuities or kinks, which would otherwise be expected by a shift by a polarization quantum \cite{spaldin2012beginner}. It naturally follows that this higher-order polarization is entirely electronic in nature, since the net ionic contribution is first-order.

Fig.~\ref{fig:high_dyn} demonstrates a case where the contribution of second-order dynamical charge to polarization can be comparable in magnitude to the first-order contribution. A strong trilinear interaction enabled by the second-order dipole gives rise to a novel form of ferroelectricity, which we termed the ``hybrid-triggered'' type in an earlier work \cite{jung2026triggered,zhao2025nature}. This observation raises several critical questions, such as the origin of the anomalously large second-order contribution in HfO$_2$---particularly $Z^{*(2)}_{x,\mathrm{O1}y,\mathrm{O1}z}$---and why similar behavior has not been reported in other materials. 
We previously identified a bond-to-bond charge transfer mechanism, wherein each nonpolar mode is associated with a bidirectional redistribution of charge between HfO$_2$ bonds. Starting from a symmetric charge distribution, this bidirectional transfer does not yield a net polar state. However, when a pre-existing charge imbalance is introduced by another nonpolar mode, the bidirectional transfer no longer cancels out. In such cases, charge transfer involving electron-rich bonds outweighs that from electron-deficient bonds, thereby producing a net polarization.

\begin{table}
\caption{\label{tab:charges} 
Second-order mode effective charges $Z^{*(2)}_{z,\lambda\mu}$ of fluorite MO$_2$ (M = Hf, Zr, Ti) and perovskite ATiO$_3$ (A = Sr, Ba, Pb). Units are in $|e|/$Å}
\begin{ruledtabular}
\begin{tabular}{ccccc}
  & $\mathrm{X}^+_{5,y}\mathrm{X}^-_{3,y}$ & $\mathrm{X}^+_{5,x}\mathrm{X}^-_{2,x}$ & $\mathrm{X}^+_{5,z}\mathrm{X}^-_{5,z:\mathrm{M}}$ & $\mathrm{X}^+_{5,z}\mathrm{X}^-_{5,z:\mathrm{O}}$ \\
\hline
HfO$_2$ 
& --1.24 & 2.72 & --1.91 & 2.11 \\
ZrO$_2$ 
& --1.38 & 3.27 & --2.46 & 2.48 \\
TiO$_2$
& --2.34 & 5.23 & --3.97 & 3.37  \\
\hline
\hline
& \multicolumn{2}{c}{$\mathrm{X}^+_{1,z:\mathrm{Ti}}\mathrm{X}^-_{3,z:\mathrm{O}}$} & \multicolumn{2}{c}{$\mathrm{X}^+_{5,z:\mathrm{O}}\mathrm{X}^-_{5,z:\mathrm{O}}$} \\
\hline
SrTiO$_3$ & \multicolumn{2}{c}{5.17} & \multicolumn{2}{c}{--0.05} \\
BaTiO$_3$ & \multicolumn{2}{c}{5.11} & \multicolumn{2}{c}{0.06} \\
PbTiO$_3$ & \multicolumn{2}{c}{5.06} & \multicolumn{2}{c}{0.07} \\
\end{tabular}
\end{ruledtabular}
\end{table}

\begin{table}
\caption{\label{tab:multipole}
Symmetry-allowed electronic multipole moments of the $\Gamma^-_4$ charge density 
at the oxygen site $(0.25,0.25,0.25)$ in space group Fm$\bar{3}$m. 
The charge density projection onto the $\Gamma^-_4$ irrep 
is obtained from the structure where hybrid nonpolar modes distortions corresponding to $Q_\lambda=7.1$ mÅ are present. Units are reported in $\upmu |e|\text{\AA}^{\ell}$.
}
\begin{ruledtabular}
\begin{tabular}{ccccc} 
 & {$\mathrm{X}^+_{5,y}\mathrm{X}^-_{3,y}$} & $\mathrm{X}^+_{5,x}\mathrm{X}^-_{2,x}$ & $\mathrm{X}^+_{5,z}\mathrm{X}^-_{5,z:\mathrm{Hf}}$ & $\mathrm{X}^+_{5,z}\mathrm{X}^-_{5,z:\mathrm{O}}$ \\
\hline
$p_z$ & --18.0 & 4.39 & --5.36 & 3.75 \\
$d_{xy}$ & 1.60 & --208 & 9.03 & --205 \\
$f_{z^3}$ & --3.38 & --2.71 & 1.63 & -2.90 \\
$g_{xyz^2}$ & --1.05 & 2.17 & 0.97 & 0.47\\
$h_{z^5}$ & 0.81 & 0.77 & 0.29 & 1.02 \\
$h_{z(x^4-6x^2y^2+y^4)}$ & --0.46 & --3.28 & 0.51 & --3.33 \\
\end{tabular}
\end{ruledtabular}
\end{table}

\section{origin of higher-order dynamical charges from Electron density projections}

To gain deeper insight into the higher-order dynamical charge contributions, we utilize projection operators \cite{Buiarelli2025ProDenCer, buiarelli2025noncollinear},  isolating the component 
of the charge density of a distorted structure that transforms according to the three-dimensional $\Gamma^-_4$ irrep of the parent space group, which is the polar representation. The isosurfaces of charge density from DFT after this projection, where crystal structures with the amplitudes of each individual mode are fixed to 7.1 mÅ, are presented in Fig.~\ref{fig:charge_projection}. Yellow and cyan lobes indicate increase and decrease in the charge densities, respectively. The projected charge densities corresponding to the two combinations of nonpolar modes observed in ferroelectric Pca2$_1$ (\#29) HfO$_2$, namely $\mathrm{X}^+_{5,z}\mathrm{X}^-_{5,z:\mathrm{Hf}}$, and $\mathrm{X}^+_{5,x}\mathrm{X}^-_{2,x}$, are shown in panels (a--b). The associated second-order mode effective charges are listed in Table.~\ref{tab:charges}, including other mode pairs $\mathrm{X}^+_{5,y}\mathrm{X}^-_{2,y}$ and $\mathrm{X}^+_{5,z}\mathrm{X}^-_{5,z:\mathrm{O}}$.

The projected charge distributions corresponding to hybrid mode $\mathrm{X}^+_{5,z}\mathrm{X}^-_{5,z:\mathrm{Hf}}$, shown in panel (a), reveal a net upward redistribution of electrons localized around both Hf and O atoms, yielding an electric dipole directed along $-z$. In other words, even though neither $\mathrm{X}^+_{5,z}$ nor $\mathrm{X}^-_{5,z:\mathrm{Hf}}$ modes are polar, their combination give rise to parallel atomic dipoles on each ion. The lobes around oxygen atoms are strongly aligned with the tetrahedral Hf–O bonds, highlighting a bond-to-bond charge transfer mechanism that scales with the amplitudes of both $\mathrm{X}^+_{5,z}$ and $\mathrm{X}^-_{5,z:\mathrm{Hf}}$. Moreover, the projected charge density reveals that a significant portion of the dipole originates from the Hf ion, most likely due to electrostatic interactions with the induced charge arising from this bond-to-bond transfer. The second-order polarization from all sites are well aligned, resembling a ferroic order.

The projected electron density of $\mathrm{X}^+_{5,x}\mathrm{X}^-_{2,x}$ hybrid mode, displayed in panel (b) is less informative due to large quadrupole contributions. To mitigate this, we employ a multipole expansion of the projected charge densities at an oxygen site. In other words, we subsequently project the local electron densities onto the basis functions of the polar irrep of the site symmetry group of an oxygen ion ($\mathrm{T}_2$). 
The set of symmetry-allowed tesseral harmonic functions for $\mathrm{T}_2$ irrep of point group $\bar{4}$3m (the site symmetry the Wyckoff position $8c$ of space group Fm$\bar{3}$m \cite{elcoro2017double}), are summarized in Table~\ref{tab:multipole}.
The combined effect of inversion-odd multipole terms (dipole, octupole, 32-pole, etc.) is illustrated in Figs.~\ref{fig:charge_projection}(c--d), corresponding to the hybrid modes 
$\mathrm{X}^+_{5,z}\mathrm{X}^-_{5,z:\mathrm{Hf}}$ and 
$\mathrm{X}^+_{5,x}\mathrm{X}^-_{2,x}$, respectively. 
The inversion-odd components of the 
$\mathrm{X}^+_{5,x}\mathrm{X}^-_{2,x}$ projected charge density indicate signatures of bond-to-bond charge transfer similar to $\mathrm{X}^+_{5,z}\mathrm{X}^-_{5,z:\mathrm{Hf}}$ with separated electron density lobes oriented along the Hf–O bonds. Projection onto multipoles further confirms that the second-order polarization aligns at all sites for the studied fluorite mode couplings.

\section{higher-order dynamical charges in other materials}

The anomalously large BECs in titanate perovskites originate from dynamical changes in the Ti–O hybridization \cite{ghosez1995born}. Although the Hf ion is more electropositive than Ti and thus forms a predominantly ionic Hf–O bond, the bond-to-bond charge-transfer mechanism in HfO$_2$ is governed primarily by covalency variations induced by changes in bond length, rather than by the degree of covalency itself \cite{yang2024origin}. Such sensitivity to covalency is also expected to become more pronounced when the underlying covalency is already substantial. Consistent with this picture, a systematic increase in the second-order mode effective charges is observed across fluorite MO$_2$ compounds as the cation electronegativity increases from Hf to Ti (Table~\ref{tab:charges}) \cite{zheng2005electronic}, further supporting the interpretation of second-order dynamical charges as manifestations of covalency-driven bond-to-bond charge transfer.

However, the observed trend of increasing second-order dynamical charge raises an important question. Why is this effect absent in the majority of Ti-based oxides, which have been extensively studied? For example, in the Ruddlesden-Popper phase A$_3$B$_2$O$_7$, the well-known hybrid mode formed by $\mathrm{X}^-_2$ and $\mathrm{X}^+_3$ transforms like the polar $\Gamma^-_5$ mode and underlies hybrid improper ferroelectricity \cite{benedek2011hybrid,benedek2012polar,oh2015experimental,li2020}. In practice, however, the polarization arises almost entirely from the first-order Born effective charges of the polar mode \cite{jung2026triggered}. Also, a recent report shows that the trilinear and quadlinear couplings which involves the $\Gamma^-_4$ mode play an important role for Pna$2_1$ phase of perovskite structures \cite{scott2024universal}. As identified here, $\mathrm{X}^+_5\mathrm{X}^-_5$ coupling transform like $\Gamma^-_4$, which means that it would enable second-order mode effective charge. 

\begin{figure}[t]
\includegraphics[width=0.47\textwidth]{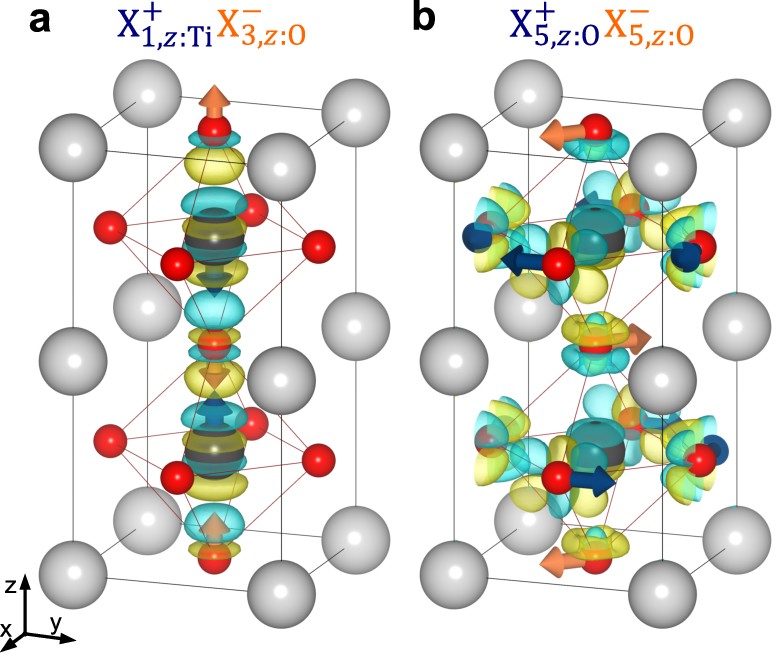}
\caption{ \label{fig:STO}
    Hybrid modes (a) $\mathrm{X}^+_{1,z:\mathrm{Ti}}\mathrm{X}^-_{3,z:\mathrm{O}}$ and (b) $\mathrm{X}^+_{5,z:\mathrm{O}}\mathrm{X}^-_{5,z:\mathrm{O}}$ in SrTiO$_3$, and the electron density projections onto the $\Gamma_4^-$ irrep. The order parameter for each mode is fixed at 0.05 Å, with isosurface values set to 1.0 m$|e|$/Å$^3$ and 0.013 m$|e|$/Å$^3$ for panels (a) and (b), respectively. The second-order polarization exhibits strong alignment at each site in panel (a), whereas this alignment is absent in panel (b).
}
\end{figure}

The reason these modes do not contribute significantly to the dipole is because the nonpolar modes in these systems mostly consist of O displacements perpendicular to the existing Ti–O bonds, such as the octahedral rotations. Such displacements produce only second‐order modifications to the bond lengths, in contrast to the relevant modes in HfO$_2$, where ions move directly toward one another leading to first‐order changes in bond lengths with respect to the order parameter.

Indeed, our first-principles calculations confirm that hybrid modes with significant second-order mode effective charges exist even in well-studied Ti-based oxides. In Fig.~\ref{fig:STO}(a), we show two zone boundary modes that are present in cubic perovskites with space group  Pm$\bar{3}$m (\#221), which we label $\mathrm{X}^+_{1,z:\mathrm{Ti}}$ and $\mathrm{X}^-_{3,z:\mathrm{O}}$. We find that in SrTiO$_3$, the second-order mode effective charge $Z^{*(2)}_{z,\mathrm{X}^+_{1,z:\mathrm{Ti}}\mathrm{X}^-_{3,z:\mathrm{O}}}$ is as large as 5.17 $|e|$/Å, comparable to the largest second-order mode effective charge observed in fluorite TiO$_2$.  On the other hand, another hybrid mode pair formed by the $\mathrm{X}^+_{5,z:\mathrm{O}}$ and
$\mathrm{X}^-_{5,z:\mathrm{O}}$ modes (Fig.~\ref{fig:STO}(b)), which only involve oxygen ions moving normal to Ti-O bonds, has the mode effective charge $Z^{*(2)}_{z,\mathrm{X}^+_{5,z:\mathrm{O}}\mathrm{X}^+_{5,z:\mathrm{O}}}$ of --0.05 $|e|$/Å. Repeating the calculation for BaTiO$_3$ and PbTiO$_3$, together with charge density projection to the $\Gamma_4^-$ irrep, reveal there is always strong alignment of electronic dipole resulting from Ti-O bond variation with the $\mathrm{X}^+_{1,z:\mathrm{Ti}}\mathrm{X}^-_{3,z:\mathrm{O}}$ modes, whereas the dipoles resulting from $\mathrm{X}^+_{5,z:\mathrm{O}}\mathrm{X}^+_{5,z:\mathrm{O}}$ modes don't have a clear, consistent alignment at each site. In a hypothetical phase where these modes condense in a ferroelectric, the nature of their contribution to the polarization would be very different, and not every pair of hybrid modes that are polar lead to a significant contribution to the switchable, \textit{ferroic} polarization.

It is worth emphasizing that in the ferroelectric Pca2$_1$ phase of HfO$_2$, all dipoles arising from second-order mode effective charges are aligned opposite to the first-order dipoles \cite{jung2026triggered}. This leads to an interesting interplay between ionic and electronic polarization effects that have possible implications for depolarization and finite size effects \cite{junquera2003critical,jung2023programmable,cheema2020enhanced}.

\section{conclusions}

In summary, by using first-principles calculations, we showed that combinations of two nonpolar structural modes can give rise to large electronic dipole moments, which can be understood by expansion of the Born effective charges framework to higher order. The fluorite structure, with many different modes that condense in the polar phase of HfO$_2$, provides an example of a system where the higher order contributions are essential to understand the magnitude of the polarization. We showed that pairs of modes that strongly modify the transition metal-oxygen bond lengths give rise to large second-order charges even in well studied perovskite systems, which indicates that the effects of higher order dynamical charges should be more prevalent than hitherto assumed. This phenomenon can influence a broad spectrum of dipole-dependent material properties, including polarization-field hysteresis, ferroelectric size effects, and piezoelectricity.

\begin{acknowledgments}
\textit{Acknowledgements}---We acknowledge helpful discussions and comments from Prof.~Nicola A. Spaldin (ETH Z\"{u}rich), Dr.~Nives Strkalj (Institute of Physics, Zagreb), Prof.~David Vanderbilt (Rutgers University) and Prof.~Philippe Ghosez (University of Li\`ege). We acknowledge the Minnesota Supercomputing Institute (MSI) for providing resources that contributed to the research results. This work is supported by the Office of Naval Research Grant N00014-24-1-2082. 
\end{acknowledgments}

\textit{Data Availability}---The data that support the findings of this article are openly available  \cite{prodencerGitHub}.

\bibliography{references}
\end{document}


\title{
\textbf{Ferroic Polarization from Nonpolar Phonons}\\[3ex] \normalsize \textsl{Supplementary Material}
}

\author{Seongjoo Jung}
\affiliation{Department of Chemical Engineering and Materials Science, University of Minnesota, Minneapolis, Minnesota 55455, USA}
\affiliation{Department of Materials, ETH Z\"{u}rich, CH-8093, Switzerland}
\author{Turan Birol}
\email{Corresponding author: tbirol@umn.edu}
\affiliation{Department of Chemical Engineering and Materials Science, University of Minnesota, Minneapolis, Minnesota 55455, USA}

\maketitle

\vspace{-0.8cm}

\section{Methods}

Periodic density functional theory (DFT) calculations were performed using the Vienna Ab Initio Simulation Package (VASP) 6.4.1 \cite{kresse1996efficient}. Spin-polarization tests confirmed that all systems considered are non-magnetic. Exchange–correlation effects were treated within the generalized gradient approximation (GGA) using the PBEsol functional \cite{perdew2008restoring}. Interactions between ionic cores and valence electrons were described using the projector augmented-wave (PAW) method. A plane-wave kinetic energy cutoff of 520 eV was employed for the Kohn–Sham orbitals. The following valence electron configurations were used: Hf ($5p^65d^36s^1$), O ($2s^22p^4$), Zr ($4s^24p^64d^35s^1$), Ti ($3s^23p^63d^34s^1$), Sr ($4s^24p^64d^{0.001}5s^{1.999}$), Ba ($5s^2 5p^6 5d^{0.01} 6s^{1.99}$), and Pb ($5d^{10} 6s^2 6p^2$).

Self consistent electronic structural calculations were converged to within $10^{-7}$ eV in electronic energy and 0.001 eV/\AA~ was the convergence threshold for residual ionic forces in structural relaxations. Born effective charges were calculated using density functional perturbation theory (DFPT) \cite{gajdovs2006linear} with a stricter electronic convergence criterion of $10^{-9}$ eV. Tetrahedron smearing with Bl\"ochl corrections was applied for Brillouin-zone integrations.

The cubic conventional cell lattice parameters ($N=4$) were 5.02 Å for HfO$_2$, 5.07 Å for ZrO$_2$, and 4.78 Å for the hypothetical structure of zero-pressure fluorite TiO$_2$.
A $\Gamma$-centered $7 \times 7 \times 7$ \textbf{k}-point mesh was used for structural relaxation, and a $9 \times 9 \times 9$ mesh was used for Born effective charge and charge density calculations. Charge densities were represented on a $120 \times 120 \times 120$ real-space grid. Convergence of the second-order dynamical charge $Z^{*(2)}$ was assessed using a $2 \times 2 \times 2$ supercell ($N=32$) and a $\Gamma$-centered $5 \times 5 \times 5$ \textbf{k}-point mesh. For SrTiO$_3$, the cubic lattice parameter was 3.86 Å. A 10-atom supercell ($N=2$) was employed to represent the $\mathrm{X}$-point modes, sampled using a $9 \times 9 \times 5$ \textbf{k}-point mesh.

Space-group representations, mode definitions, and symmetry couplings were obtained from the Bilbao Crystallographic Server \cite{aroyo2011crystallography} and from FINDSYM, SMODES, and INVARIANTS in the ISOTROPY software suite \cite{hatch2003invariants,stokes2005findsym,stokes_isotropy}. Order parameters follow the normalization convention which includes a factor of $\sqrt{N}$ based on the number of primitive cells \cite{cowley1980structural}:
\begin{gather}
    Q_\lambda \frac{\bm{U}_\lambda}{|\bm{U}_\lambda|}= \frac{\bm{u}_\lambda}{\sqrt{N}} \label{eq:mode_def} \\
    |Q_\lambda| = \frac{|\bm{u}_\lambda|}{\sqrt{N}} = 
    \sqrt{\frac{\sum_{\kappa\alpha} u_{\lambda,\kappa\alpha}^2}{N}}
\end{gather}
The second-order mode effective charges are calculated using finite differences from eq.~(3), using multiple absolute values of order parameters between 0.01-0.1 Å.
Multipole moments were computed using the following definition, with a cutoff radius $R = 1.20$ Å:
\begin{gather}
    w_{\ell m} = \int_{|\mathbf{r}-\mathbf{r}_0|<R} 
    \sqrt{\frac{4\pi}{2\ell+1}} \, r^\ell \, \rho(\mathbf{r}) \, 
    T_{\ell m}(\mathbf{r})\, d^3\mathbf{r}
\end{gather}
where $w_{\ell m}$ are the electric multipole moments, $\rho$ is the charge density (negative for positive electron density), $T_{\ell m}$ are tesseral harmonic functions, $\ell$ is the orbital angular momentum quantum number, and $m$ is the magnetic quantum number. 
Crystal structures were constructed and visualized using VESTA \cite{momma2011vesta}.

\section{Second-order Mode Effective Charges}

The dipole moment in a unit cell can be expanded up to second order with respect to ionic sublattice displacements:
\begin{equation}
 p_\alpha = \Omega P_\alpha = \sum_{\kappa \beta} Z^{*}_{\alpha,\kappa\beta}u_{\kappa \beta} + \frac{1}{2}\sum_{\kappa' \gamma} \sum_{\kappa \beta} Z^{*(2)}_{\alpha,\kappa\beta,\kappa'\gamma}u_{\kappa \beta}u_{\kappa' \gamma}   
\end{equation}
Considering the situation where only two nonpolar modes $\lambda$ and $\mu$ are present, the second-order mode effective charge for these modes can be calculated by resolving each displacement $u_{\kappa \beta}$ to the modes' displacements $u_{\lambda,\kappa \beta}$ and $u_{\mu,\kappa \beta}$:
\begin{widetext}
\begin{equation}
\begin{aligned}
 p_\alpha &= \frac{1}{2}\sum_{\kappa' \gamma} \sum_{\kappa \beta} Z^{*(2)}_{\alpha,\kappa\beta,\kappa'\gamma}(u_{\lambda,\kappa \beta}+u_{\mu,\kappa \beta})(u_{\lambda,\kappa' \gamma}+u_{\mu,\kappa' \gamma})   \\
  &=   \frac{1}{2} \left( \sum_{\kappa' \gamma} \sum_{\kappa \beta} Z^{*(2)}_{\alpha,\kappa\beta,\kappa'\gamma}u_{\lambda,\kappa \beta}u_{\lambda,\kappa' \gamma} 
  + \sum_{\kappa' \gamma} \sum_{\kappa \beta} Z^{*(2)}_{\alpha,\kappa\beta,\kappa'\gamma}u_{\lambda,\kappa \beta}u_{\mu,\kappa' \gamma} 
  + \sum_{\kappa' \gamma} \sum_{\kappa \beta} Z^{*(2)}_{\alpha,\kappa\beta,\kappa'\gamma}u_{\mu,\kappa \beta}u_{\lambda,\kappa' \gamma}
  + \sum_{\kappa' \gamma} \sum_{\kappa \beta} Z^{*(2)}_{\alpha,\kappa\beta,\kappa'\gamma}u_{\mu,\kappa \beta}u_{\mu,\kappa' \gamma} \right)
\end{aligned}
\end{equation}
\end{widetext}
From eq.~(2) and (\ref{eq:mode_def}), the above equation reduces to:
\begin{align}
 p_\alpha = \frac{N}{2}(Z^{*(2)}_{\alpha,\lambda\lambda}{Q_\lambda}^2+Z^{*(2)}_{\alpha,\lambda\mu}{Q_\lambda}{Q_\mu}+Z^{*(2)}_{\alpha,\mu\lambda}{Q_\mu}{Q_\lambda}+Z^{*(2)}_{\alpha,\mu\mu}{Q_\mu}^2)
\end{align}
Since both modes $\lambda$ and $\mu$ are nonpolar, $Z^{*(2)}_{\lambda\lambda}=Z^{*(2)}_{\mu\mu}=0$
and from eq.~(2) $Z^{*(2)}_{\lambda\mu}=Z^{*(2)}_{\mu\lambda}$. This further reduces above equation to eq.~(3).
The factor of $N$ results simply from our treatment of order parameter $Q$ to be independent of system size. It does not appear for the choice of order parameter $Q$ from mode amplitude without normalization factor,
where the value of $Q$ would be proportional to $\sqrt{N}$. Regardless of this, the value of second-order mode effective charge following the definition of eq.~(2) does not depend on the choice of unit cell.

\section{Unit cell dependence of Born effective charge and second-order dynamical charge}

The definition of the Born effective charge involves the displacement of sublattice ions and therefore may appear to depend on the choice of unit cell under periodic boundary conditions:
\begin{align}
Z^*_{\alpha,\kappa\beta} = \Omega\,\pdv{P_\alpha}{u_{\kappa\beta}} .
\end{align}
However, one can show that the Born effective charge itself is invariant with respect to the unit cell choice in linear order. Consider a situation in which the sublattice $\kappa$ is partitioned into $\kappa_1$ and $\kappa_2$ when the unit cell is doubled from $C_1$ to $C_2$. In $C_1$, the linear response of the dipole moment $\Omega P_\alpha$ to a small displacement $u_{\kappa\beta}$ is $Z^*_{\alpha,\kappa\beta}\,u_{\kappa\beta}$. For the same displacement $u_{\kappa\beta}$ in $C_2$ (so that $u_{\kappa\beta}=u_{\kappa_1\beta}=u_{\kappa_2\beta}$), the dipole moment becomes
\begin{align}
2Z^*_{\alpha,\kappa\beta}u_{\kappa\beta}
  = Z^*_{\alpha,\kappa_1\beta}u_{\kappa_1\beta}
  + Z^*_{\alpha,\kappa_2\beta}u_{\kappa_2\beta}.
\end{align}
Because $\kappa_1$ and $\kappa_2$ are related by the translational symmetry of $C_1$, one has
$Z^*_{\alpha,\kappa_1\beta}=Z^*_{\alpha,\kappa_2\beta}$, and therefore
$Z^*_{\alpha,\kappa\beta}=Z^*_{\alpha,\kappa_1\beta}=Z^*_{\alpha,\kappa_2\beta}$.

In contrast, the second-order dynamical charges of ion pairs depend on the unit cell choice, since they are inherently pairwise quantities. The total contribution to the electronic dipole $\Omega P_\alpha$ from the second-order dynamical charges is
\begin{align}
\frac{1}{2}\sum_{\kappa\beta}\sum_{\kappa'\gamma}
  Z^{*(2)}_{\alpha,\kappa\beta,\kappa'\gamma}\,
  u_{\kappa\beta}u_{\kappa'\gamma}.
\end{align}
Again consider doubling the unit cell so that $\kappa\to\{\kappa_1,\kappa_2\}$ and
$\kappa'\to\{\kappa'_1,\kappa'_2\}$. In $C_1$, with displacements
$u_{\kappa\beta}$ and $u_{\kappa'\gamma}$ (Here, $\kappa, \kappa', \beta$ and $\gamma$ refer to specific indices instead of generic ones), the dipole moment from the second-order term is
$Z^{*(2)}_{\alpha,\kappa\beta,\kappa'\gamma}u_{\kappa\beta}u_{\kappa'\gamma}$, assuming
$Z^{*(2)}_{\alpha,\kappa\beta,\kappa\beta}
 =Z^{*(2)}_{\alpha,\kappa'\gamma,\kappa'\gamma}=0$, which holds for most high-symmetry sites.
In $C_2$, one obtains
\begin{align}
2Z^{*(2)}_{\alpha,\kappa\beta,\kappa'\gamma}u_{\kappa\beta}u_{\kappa'\gamma}
  &=
  2Z^{*(2)}_{\alpha,\kappa_1\beta,\kappa'_1\gamma}
   u_{\kappa_1\beta}u_{\kappa'_1\gamma}
  +2Z^{*(2)}_{\alpha,\kappa_1\beta,\kappa'_2\gamma}
   u_{\kappa_1\beta}u_{\kappa'_2\gamma}
  \nonumber\\
  &\quad
  +Z^{*(2)}_{\alpha,\kappa_1\beta,\kappa_2\beta}
   u_{\kappa_1\beta}u_{\kappa_2\beta}
  +Z^{*(2)}_{\alpha,\kappa'_1\gamma,\kappa'_2\gamma}
   u_{\kappa'_1\gamma}u_{\kappa'_2\gamma},
\end{align}
after applying the translational symmetry of $C_1$. Thus, in general,
$Z^{*(2)}_{\alpha,\kappa\beta,\kappa'\gamma}
 \neq Z^{*(2)}_{\alpha,\kappa_1\beta,\kappa'_1\gamma}$.

The following section presents the second-order dynamical charges of HfO$_2$ in an $N=32$ supercell, for which the unit cell has been octupled relative to Fig.~1(b).

\clearpage
\newpage

\section{Length Scale of higher-order dynamical charges}

\begin{figure}[!htb]
    \centering
    \includegraphics[width=0.76\linewidth]{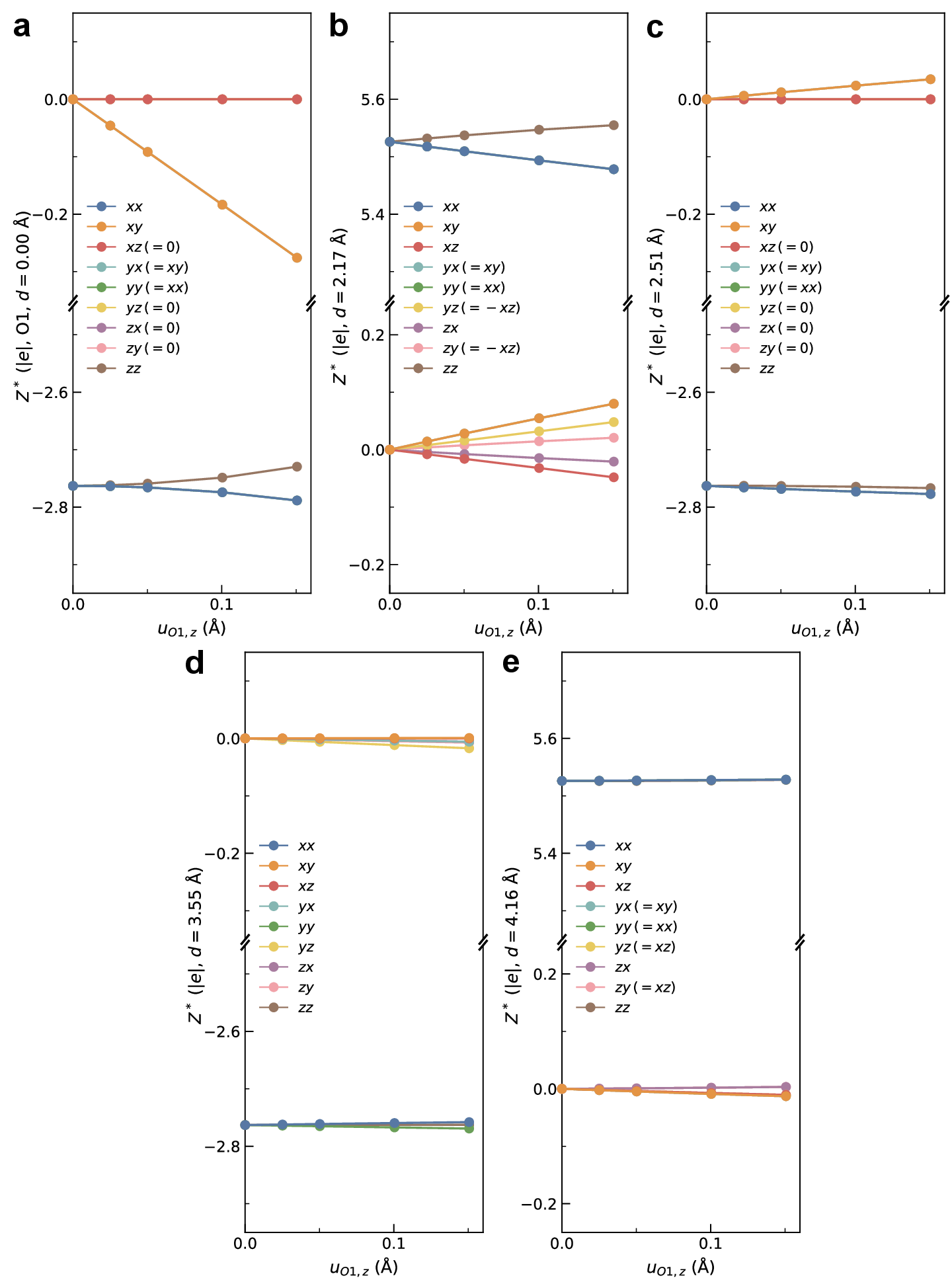}
    \caption{
    Change of Born effective charge components by Oxygen ion  displacement $u_{\mathrm{O1}z}$ in cubic HfO$_2$ supercell ($N=32$) in 
    (a) itself, O at (0.5, 0.5, 0.5) in direct coordinates
    (b) first nearest neighbor, Hf at (0.375, 0.625, 0.375)
    (c) second nearest neighbor, O at (0.5, 0.5, 0.25)
    (d) third nearest neighbor, O at (0.25, 0.5, 0.25)
    (e) fourth nearest neighbor, Hf at (0.375, 0.375, 0.125)
    }
    \label{sup:BEC}
\end{figure}

Fig.~\ref{sup:BEC} shows the change of BEC components in response to $z$ displacement of an oxygen at direct coordinates (0.5, 0.5, 0.5) for the $N=32$ supercell of HfO$_2$ under periodic boundary conditions. Difference in sign from Fig.~1(b) results from different bonding environment of the center oxygen ion. The difference in $Z^{*(2)}_{x,\mathrm{O1}y,\mathrm{O1}z}$ is less than 10\% from the $N=4$ supercell, and minimal change in BEC components are observed for ion pairs further than second nearest neighbor ($d\sim3$ Å). Note that in the labels, the first index is for direction of polarization, and the second index is for direction of sublattice displacement.

\newpage
\section{Symmetry-Mode definitions}

In this section, we provide a list of normalized symmetry-mode displacements for different modes that are referred to in the main text and the Supplementary Material. 

\begin{table}[!htp]
\centering
\renewcommand{\arraystretch}{1.3}
\caption{
Direct coordinates of ions in cubic HfO$_2$ conventional unit cell, $N=4$.
}
\begin{tabularx}{0.4\textwidth}{Y|YYY}
\hline\hline
Ion & $x$ & $y$ & $z$ \\
\hline
Hf1 & 0 & 0 & 0 \\
Hf2 & 0.5 & 0.5 & 0 \\
Hf3 & 0.5 & 0 & 0.5 \\
Hf4 & 0 & 0.5 & 0.5 \\
O1 & 0.25 & 0.25 & 0.25 \\
O2 & 0.75 & 0.25 & 0.25 \\
O3 & 0.25 & 0.75 & 0.25 \\
O4 & 0.75 & 0.75 & 0.25 \\
O5 & 0.25 & 0.25 & 0.75 \\
O6 & 0.75 & 0.25 & 0.75 \\
O7 & 0.25 & 0.75 & 0.75 \\
O8 & 0.75 & 0.75 & 0.75 \\
\hline\hline
\end{tabularx}
\end{table}

\begin{table}[!htp]
\caption{
Symmetry-mode definitions (${\bm{U}_\lambda}/{|\bm{U}_\lambda|}$) in HfO$_2$ conventional unit cell, $N=4$.
}
\renewcommand{\arraystretch}{1.3}
\begin{tabularx}{0.5\textwidth}{Y|Y|YYY}
\hline\hline
Mode & Ion & $x$ & $y$ & $z$ \\
\hline
\multirow{8}{*}{$\Gamma_{4}^-$}  & O1 & 0 & 0 & $-0.3536$ \\
                                 & O2 & 0 & 0 & $-0.3536$ \\
                                 & O3 & 0 & 0 & $-0.3536$ \\
                                 & O4 & 0 & 0 & $-0.3536$ \\
                                 & O5 & 0 & 0 & $-0.3536$ \\
                                 & O6 & 0 & 0 & $-0.3536$ \\
                                 & O7 & 0 & 0 & $-0.3536$ \\
                                 & O8 & 0 & 0 & $-0.3536$ \\
\hline
\multirow{8}{*}{$\mathrm{X}_{5,y}^+$} & O1 & 0 & 0 & $+0.3536$ \\
                                 & O2 & 0 & 0 & $+0.3536$ \\
                                 & O3 & 0 & 0 & $-0.3536$ \\
                                 & O4 & 0 & 0 & $-0.3536$ \\
                                 & O5 & 0 & 0 & $+0.3536$ \\
                                 & O6 & 0 & 0 & $+0.3536$ \\
                                 & O7 & 0 & 0 & $-0.3536$ \\
                                 & O8 & 0 & 0 & $-0.3536$ \\
\hline
\multirow{8}{*}{$\mathrm{X}_{5,x}^+$} & O1 & 0 & $+0.3536$ & 0 \\
                                 & O2 & 0 & $-0.3536$ & 0 \\
                                 & O3 & 0 & $+0.3536$ & 0 \\
                                 & O4 & 0 & $-0.3536$ & 0 \\
                                 & O5 & 0 & $+0.3536$ & 0 \\
                                 & O6 & 0 & $-0.3536$ & 0 \\
                                 & O7 & 0 & $+0.3536$ & 0 \\
                                 & O8 & 0 & $-0.3536$ & 0 \\
\hline
\hline
\end{tabularx}
\end{table}

\setcounter{table}{1}
\begin{table}[!htp]
\caption{
(Continued)
}
\renewcommand{\arraystretch}{1.3}
\begin{tabularx}{0.5\textwidth}{Y|Y|YYY}
\hline\hline
Mode & Ion & $x$ & $y$ & $z$ \\
\hline
\multirow{8}{*}{$\mathrm{X}_{5,z}^+$} & O1 & $+0.3536$ & 0 & 0 \\
                                 & O2 & $+0.3536$ & 0 & 0 \\
                                 & O3 & $+0.3536$ & 0 & 0 \\
                                 & O4 & $+0.3536$ & 0 & 0 \\
                                 & O5 & $-0.3536$ & 0 & 0 \\
                                 & O6 & $-0.3536$ & 0 & 0 \\
                                 & O7 & $-0.3536$ & 0 & 0 \\
                                 & O8 & $-0.3536$ & 0 & 0 \\
\hline
\multirow{8}{*}{$\mathrm{X}_{2,x}^-$} & O1 & $+0.3536$ & 0 & 0 \\
                                 & O2 & $+0.3536$ & 0 & 0 \\
                                 & O3 & $-0.3536$ & 0 & 0 \\
                                 & O4 & $-0.3536$ & 0 & 0 \\
                                 & O5 & $-0.3536$ & 0 & 0 \\
                                 & O6 & $-0.3536$ & 0 & 0 \\
                                 & O7 & $+0.3536$ & 0 & 0 \\
                                 & O8 & $+0.3536$ & 0 & 0 \\
\hline
\multirow{4}{*}{$\mathrm{X}_{3,y}^-$} & Hf1 & 0 & $+0.5$ & 0 \\
                                 & Hf2 & 0 & $-0.5$ & 0 \\
                                 & Hf3 & 0 & $+0.5$ & 0 \\
                                 & Hf4 & 0 & $-0.5$ & 0 \\
\hline
\multirow{4}{*}{$\mathrm{X}_{5,z:\mathrm{Hf}}^-$} & Hf1 & $+0.5$ & 0 & 0 \\
                                 & Hf2 & $+0.5$ & 0 & 0 \\
                                 & Hf3 & $-0.5$ & 0 & 0 \\
                                 & Hf4 & $-0.5$ & 0 & 0 \\
\hline
\multirow{8}{*}{$\mathrm{X}_{5,z:\mathrm{O}}^-$} & O1 & 0 & $+0.3536$ & 0 \\
                                 & O2 & 0 & $-0.3536$ & 0 \\
                                 & O3 & 0 & $-0.3536$ & 0 \\
                                 & O4 & 0 & $+0.3536$ & 0 \\
                                 & O5 & 0 & $+0.3536$ & 0 \\
                                 & O6 & 0 & $-0.3536$ & 0 \\
                                 & O7 & 0 & $-0.3536$ & 0 \\
                                 & O8 & 0 & $+0.3536$ & 0 \\
\hline
\hline
\end{tabularx}
\end{table}

\begin{table}[!htp]
\centering
\renewcommand{\arraystretch}{1.3}
\caption{
Direct coordinates of ions in cubic SrTiO$_3$ supercell, $N=2$.
}
\begin{tabularx}{0.4\textwidth}{Y|YYY}
\hline\hline
Ion & $x$ & $y$ & $z$ \\
\hline
Sr1 & 0 & 0 & 0 \\
Sr2 & 0 & 0 & 0.5 \\
Ti1 & 0.5 & 0.5 & 0.25 \\
Ti2 & 0.5 & 0.5 & 0.75 \\
O1 & 0.5 & 0.5 & 0 \\
O2 & 0.0 & 0.5 & 0.25 \\
O3 & 0.5 & 0.0 & 0.25 \\
O4 & 0.5 & 0.5 & 0.5 \\
O5 & 0.0 & 0.5 & 0.75 \\
O6 & 0.5 & 0.0 & 0.75 \\
\hline\hline
\end{tabularx}
\end{table}

\begin{table}[!htp]
\caption{
Symmetry-mode definitions (${\bm{U}_\lambda}/{|\bm{U}_\lambda|}$) in SrTiO$_3$ supercell, $N=2$.
}
\renewcommand{\arraystretch}{1.3}
\begin{tabularx}{0.5\textwidth}{Y|Y|YYY}
\hline\hline
Mode & Ion & $x$ & $y$ & $z$ \\
\hline
\multirow{2}{*}{$\mathrm{X}^+_{1,z:\mathrm{Ti}}$} & Ti1 & 0 & 0 & $+0.7071$ \\
                                 & Ti2 & 0 & 0 & $-0.7071$ \\
\hline
\multirow{2}{*}{$\mathrm{X}^-_{3,z:\mathrm{O}}$}  & O1 & 0 & 0 & $+0.7071$ \\
                                 & O4 & 0 & 0 & $-0.7071$ \\
\hline
\multirow{3}{*}{$\mathrm{X}^+_{5,z:\mathrm{O}}$}  & O2 & 0 & $+0.5$ & 0 \\
                                 & O3 & $-0.5$ & 0 & 0 \\
                                 & O5 & 0 & $-0.5$ & 0 \\
                                 & O6 & $+0.5$ & 0 & 0 \\
\hline
\multirow{2}{*}{$\mathrm{X}^-_{5,z:\mathrm{O}}$}  & O1 & $+0.5$ & $-0.5$ & 0 \\
                                 & O4 & $-0.5$ & $+0.5$ & 0 \\
\hline
\multirow{2}{*}{$\Gamma^-_{4:\mathrm{Ti}}$}       & Ti1 & 0 & 0 & $+0.7071$ \\
                                 & Ti2 & 0 & 0 & $+0.7071$ \\
\hline
\hline
\end{tabularx}
\end{table}

\clearpage
\section{$k$-grid convergence}

\begin{figure}[h]
    \centering
    \includegraphics[width=0.9\linewidth]{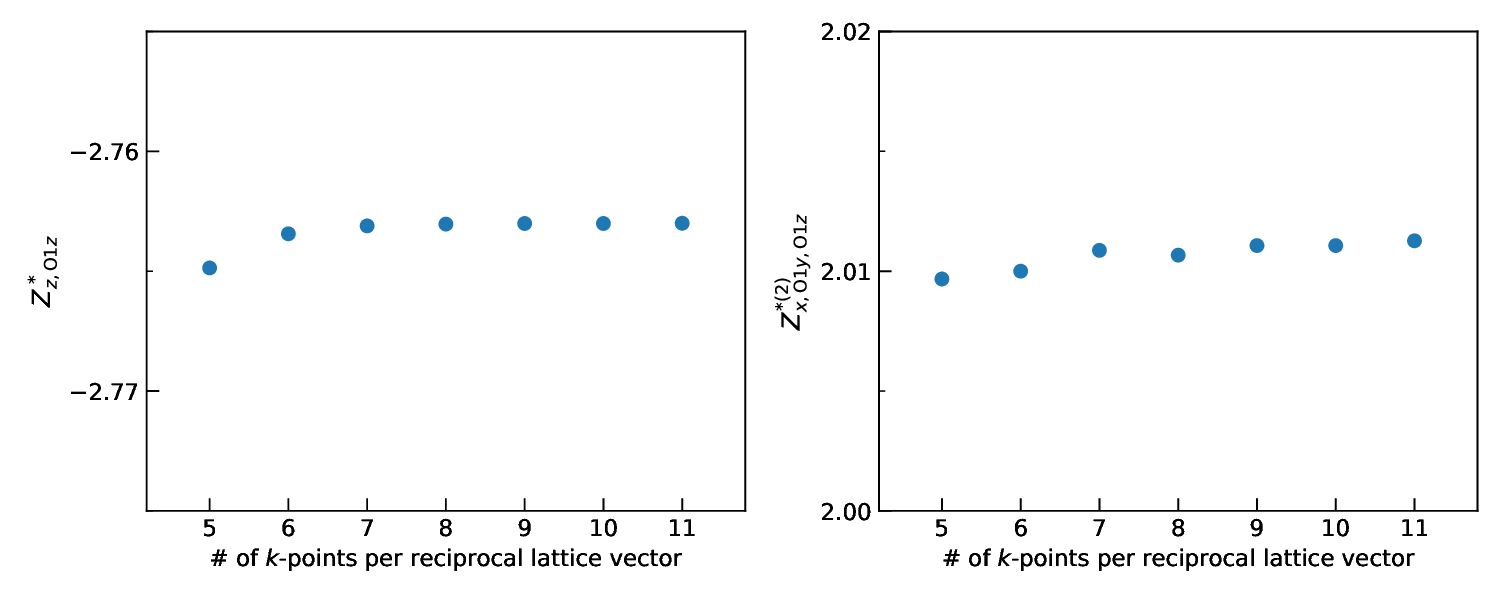}
    \caption{Convergence of $Z^*_{z,\mathrm{O1}z}$ and $Z^{*,(2)}_{x,\mathrm{O1}y,\mathrm{O1}z}$ with increasing number of points in the $k$-grid. $9 \times 9 \times 9$ grid was used for Born effective charge and second-order dynamical charge calculations.}
    \label{sup:kconv}
\end{figure}

\FloatBarrier
\bibliography{references}